# The discovery of Cherenkov radiation and its use in the detection of extensive air showers

## Alan A. Watson

## School of Physics and Astronomy, University of Leeds, Leeds LS2 9JT, UK

Cascades of charged particles are created when high-energy cosmic rays enter the earth's atmosphere: these 'extensive air-showers' are studied to gain information on the energy spectrum, arrival direction distribution and mass composition of the particles above $10^{14}$ eV where direct observations using instruments carried by balloons or satellites become impractical. Detection of light in the visible and ultra-violet ranges of the electromagnetic spectrum plays a key role in this work, the two processes involved being the emission of Cherenkov light and the production of fluorescence radiation. In this paper I will outline some of the history of the discovery of the Cherenkov process and describe the use to which it has been put in the study of extensive air-showers at ground level.

## 1  The discovery of Cherenkov radiation

The phenomenon that we now call 'Cherenkov radiation' was studied experimentally by Pavel Cherenkov and theoretically by Il'Ja Frank and Igor Tamm in the mid-1930s. Cherenkov, Frank and Tamm were awarded the Nobel Prize for this work in 1958. The experimental activity was started by Cherenkov in Leningrad in 1933 under the direction of S.I. Vavilov, the father of non-linear optics, and continued after Vavilov and his group relocated to what was to become the Lebedev Institute in Moscow the following year. The combination of elegant experimentation and sound theoretical work uncovered and explained a phenomenon that had been noted much earlier. It is evident from Eva Curie's biography of her mother that Pierre and Marie Curie, in the first years of the 20[th] Century, were very familiar with the bluish glow seen in the dark from glass vessels containing salts of radium [1]. This was undoubtedly 'Cherenkov radiation' but no systematic investigation of the phenomenon was undertaken until Mallet [2] examined the effect in some detail, noting that blue-colored light was emitted from a number of transparent objects when a radioactive source was nearby. Mallet determined the spectrum of the light (using a spectrometer designed by Fabry) establishing with photographic methods that the spectrum was continuous and extended at least to his limit of measurement at 370 nm. However Mallet took his studies no further missing the polarization of the radiation and, crucially, the asymmetry of its emission. An excellent account of the work of Mallet and of Cherenkov and of the interpretation of Cherenkov's observations by Frank and Tamm, together with extensive details about applications of the radiation, has been given by Jelley [3].

The experimental procedure adopted by Cherenkov, as proposed by Vavilov, was to irradiate various liquids with γ-rays from a 104 mg radium source. His initial task was to study fluorescence emission induced in uranyl salts. His early experiments were made using naked-eye observations with the intensity of radiation found using a technique, called 'quenching', in which the dark-adapted eye was used with a graded wedge to provide calibration of the light intensity. These were difficult and delicate experiments requiring high levels of both patience and experimental skill. Cherenkov discovered that light was emitted even when the vessel contained only sulphuric acid, the solvent for the uranyl salt. He went on to demonstrate that the light was observed in a range of different solvents. In a moving obituary [4], which mentions only briefly Cherenkov's considerable post-war contributions to accelerator physics, Chudakov writes "The phenomenon was not and probably could not have been discovered earlier by someone more experienced in physics than Cherenkov was in the 1930s. To determine the nature of the faint blue light produced in different liquids by gamma rays from a radioactive source seemed to require a young fellow from a rural area, inexperienced but with immense patience and vigour".

Observation of polarization was an important clue to the eventual interpretation but the critical breakthrough was Cherenkov's discovery in 1936 that the radiation was emitted asymmetrically only in the forward direction with respect to the direction of the incoming γ-ray beam [5], at an angle consistent with Huygens' Principle. The asymmetric emission ruled out the explanation proposed by Vavilov, namely that what was being observed was radiation from Compton electrons, created by the gamma rays, being slowed down in the liquid, that is bremsstrahlung in the optical band. It was Vavilov who asked Tamm to study the phenomenon theoretically and he and Frank, who also aided Cherenkov in some of the later observations, developed a theory in which the emission was attributed to dipole radiation arising from the transient polarisation of atoms in a medium traversed by charged particles. When such a particle moves slowly through a medium, the radiation from the excited dipoles is emitted symmetrically around the path. However, should the velocity of the particle be greater than the phase velocity of light, wavelets of electromagnetic radiation from all points along the track will be in phase and at a distance from it there will be an electromagnetic wave.

The theory is surely too well-known to readers of this note to require repetition here. The famous equation that defines the opening angle of the cone along which the radiation travels is

$$cos\ \theta = 1/\beta n \quad\quad\quad\quad\quad\quad\quad\quad\quad\quad (1)$$

where $\theta$ is the opening angle of the cone, $n$ is the refractive index of the medium and $\beta = v/c$ where $v$ is the velocity of the particle and $c$ is the speed of light. From this equation it is evident that there is a threshold value of $\beta$ below which no radiation is emitted coherently with respect to the track of the particle and that for a high-speed particle with $\beta \sim 1$, there is a maximum angle of emission, the Cherenkov angle, of $\theta_{max} = cos^{-1}(1/n)$.

Cherenkov radiation is often described as being analogous to the shock wave emitted in the acoustic range by a projectile or plane moving through air faster than the speed of sound or to the bow wave created by a boat moving across water. In recognition of this analogy Tamm [6], in his Nobel Lecture, coined the phrase 'the singing electrons', although, of course, any charged particle of appropriate speed will produce Cherenkov radiation.

In Cherenkov's first studies the charged particles were Compton electrons produced by the photons from radium. This made it difficult to check the angular dependence upon $\beta$ of equation 1 as the electrons had a range of energies. The report of the first attempt to check these predictions was published in Russian [7]: more accessible accounts on Cherenkov's work confirming the angle as function of $\beta$ are available in Jelley's book [3] and in a more recent article by Govorkov [8]. The validity of Cherenkov's observations and of equation 1 was strikingly confirmed by Collins and Resling [9] and by Wyckoff and Henderson [10] who had access to well-collimated beams of monoenergetic electrons of ~ 2 MeV.

It was Cherenkov who first suggested that the radiation might be used to detect charged particles.

## 2  Vavilov-Cherenkov radiation vs Cherenkov radiation and the recognition of Cherenkov within the USSR

In his Nobel prize address, Tamm [6] points out that "in the USSR the name 'Vavilov-Cherenkov radiation' is used instead of just 'Cherenkov radiation' in order to emphasise the role of S. Vavilov in the discovery". The role of Vavilov in the work of Cherenkov has been discussed recently by Bolotovski, Y N. Vavilov and Shmeleva [11]. Vasily Prosin has told me that, shortly before he died, G.T. Zatsepin had taken him to task for speaking of Cherenkov radiation. However in his obituary of Cherenkov, Chudakov [4] does not give Vavilov this recognition despite a delegation from the Lebedev Institute asking him to do so. I am told that Chudakov took the view that Vavilov had never fully understood the nature of the mechanism that produces the radiation.

It was pointed out by Kaiser (1973) [12] that Heaviside [13] had predicted in 1888 that an electromagnetic wave would be associated with the movement of point charge at a speed greater than that of light and Jelley [14] has suggested that entitling the radiation Heaviside-Mallet radiation would be justifiable. Sommefeld made a proposal similar to that of Heaviside but seems to have been unaware of Heaviside's insight [15]. However, what seems quite certain is that 'Cherenkov radiation' is likely to continue to be the name used world-wide when talking of the phenomenon.

Perhaps of more interest is the rather slow rate at which recognition came to Cherenkov within the USSR and why. I am conscious that, as a non-Russian and as a non-reader of Russian, my comments must be somewhat superficial but there are nonetheless some sure facts, some of which have been pointed out in a delightful article written by Cherenkov's daughter [16]. Although Cherenkov, along with Vavilov, Tamm and Frank were awarded the Stalin prize in 1946, Cherenkov's election to the Academy of Sciences USSR seems to have come unusually late. His daughter notes [16] that he became a corresponding member only in 1964, six years after the award of the Nobel Prize, and was elected as a full member 6 years later when he was 66 years old. By contrast Vavilov had become a full member in 1932 at the age of 41. Why was there this delay? Although Cherenkov's father-in-law had fallen foul of Stalin's system and been imprisoned in 1930, Stalin had died in 1953, 5 years before the award of the Nobel Prize. Vavilov was also deeply affected by the Stalinist regime: his brother, a prominent geneticist, died in prison after being sentenced to death for challenging the views of Lysenko. It may be that Cherenkov simply did not fit in well with the sophisticated lifestyle of his Moscow colleagues who would surely have had significant influence in Academy elections. He was born in a small village and educated at the University of Voronezh some 600 km from Moscow and had worked as a school teacher before going to Leningrad for post-graduate study. His background was thus rather unusual: in discussions with colleagues who have considerably more knowledge of life in the Soviet era than I have, he has been described to me as appearing like a simpleton and having peasant manners. Was there some sort of class discrimination which fueled resistance to accepting that someone from such a humble background could make such a momentous discovery? Certainly the Soviet system was a lot less equal than Westerners were encouraged to believe.

Perhaps the last words on Cherenkov's discovery are best taken from Chudakov's obituary [4]:

"Nevertheless, when considering the glorious development of the Cherenkov technique in experimental physics, I imagine a young and enthusiastic fellow who for several years started his working day by spending an hour in a totally dark room to prepare his eyes to observe faint light and who scrupulously repeated the observations again and again, varying the liquids and the geometries of the experiment, trying to find the clue to the nature of the puzzling radiation that now bears his name."

## 3  Production of Cherenkov light in air

There is no argument as to who was the first person to discuss the production of Cherenkov light in air: it was the British physicist, P.M.S. Blackett. Blackett, perhaps best known for his cloud chamber work on electromagnetic cascades carried out in Cambridge in the 1930s and which led to the Nobel Prize in 1948, also had a deep interest in the use of extensive air showers to study the highest energy cosmic rays. This interest was almost certainly stimulated by B. Rossi and P. Auger who both spent time in Blackett's laboratory in Manchester immediately prior to the start of World War II[1]. In 1948, at a meeting to discuss the Emission Spectra of the Night Sky and Aurorae [17], Blackett described the essential features of the production of Cherenkov radiation in air. He calculated that the energy thresholds (equation 1) for electrons and muons at standard temperature and pressure were 20 MeV and 4000 MeV respectively. Blackett also showed that the Cherenkov radiation produced by cosmic rays traversing the atmosphere comprised $\sim 10^{-4}$ of the night-sky background thus setting a limit to the darkness of the sky at night even under cloud. He commented that 'presumably such a small intensity of light could not be detected by normal methods'. This was Blackett's only publication on this topic but, Lovell [18] relates that he became interested in the quantum efficiency of the eye, concluding that extensive air showers should produce a flash of light that he should be able to see lying down and looking upwards under suitable dark sky conditions, an investigation which Blackett carried out himself. I do not know the outcome of this effort but his work inspired J.V. Jelley and W. Galbraith to look for flashes of light associated with extensive air showers using photomultipliers that had recently been developed rather than their eyes.

While working at the atomic energy facility at Harwell in the UK (but outside the security wire), Galbraith and Jelley[2] pointed a searchlight mirror of 25 cm diameter and ~12 cm focal length vertically with a photomultiplier of 5 cm diameter at the focus of the mirror. The output from the photomultiplier was connected to an oscilloscope. The oscilloscope was first connected directly to the photomultiplier and pulses were seen on the timebase at a rate of about 1 per minute with the threshold set at three times that of the night sky noise[3]. Following this success, the oscilloscope was triggered when Geiger counters, part of an air shower array that was being operated at Harwell, were struck in coincidence. The experiments were carried out during the moonless periods in September and October 1952 and the experiments showed that light pulses observed were associated with cosmic radiation [19]. However there was no evidence from this work to support the idea that the light observed was Cherenkov radiation, bremmsstrahlung or light from ionization being other possibilities.

---

[1] I am indebted to Professor Sir Bernard Lovell for this insight.

[2] According to Giorgio Palumbo, John Jelley was classified in the Harwell telephone directory as a 'miscellaneous researcher'.

[3] Bill Galbraith once told me that he and Jelley had been fortunate that the rate was what it was: had it been very much lower then they might well have not proceeded while had it been very much higher their task would have been more difficult.

Galbraith and Jelley continued their investigations the following year, moving to the Pic du Midi to take advantage of the greater number of nights of high clarity [20]. Four light receivers, similar to those employed at Harwell, were used in conjunction with 5 trays of Geiger counters, each containing 4 counters of 200 cm$^2$. Through an elegant series of experiments Galbraith and Jelley demonstrated that the light signals had the polarization characteristic of Cherenkov radiation and that the light had a spectral distribution consistent with what was predicted. The detection of polarization eliminated the possibility that the light seen was due to recombination radiation while calculations showed that the bremsstrahlung process did not give sufficient photons to explain the observations. In the same sequence of experiments they were able to show that there was a correlation of the light signals with shower energy: they estimated that the threshold for detection was ~$10^{14}$ eV. Galbraith and Jelley also carried out the first searches for high-energy cosmic-ray sources using the Cherenkov method to examine radio sources in the Cygnus region and in Cassiopeia as well as the Andromeda nebula and the Galactic plane. I will not discuss the history of the use of Cherenkov radiation in high-energy gamma ray astronomy as reviews are available in recent articles by Lidvansky [21] and Völk and Bernlöhr [22].

The UK does not have an optimum climate for systematic studies of showers produced by relatively low energy cosmic rays[4] and the Harwell activity was not followed up in Britain for nearly 20 years. However the field was developed vigorously during the 1950s in the USSR by Chudakov and Nesterova [23] working in the Pamirs and in Australia by Brennan and colleagues at the University of Sydney [24]. In the early 1970s, following the promptings of K. Greisen, efforts were pushed successfully in the UK by Turver and his colleagues using showers produced by primaries of energy >$10^{17}$ eV measured with the array of water-Cherenkov detectors at Haverah Park [25].

The Cherenkov light measurements in air-showers are used to deduce the energy lost by electrons through ionization, $E_i$. The light signal is strong and the total signal integrated over the shower front is closely proportional to the energy lost to ionization lost by charged particles in the atmosphere. Measurement of the Cherenkov radiation thus provides a route for finding the energy of the primary. It is also argued that the depth of shower maximum can be extracted from the lateral and temporal distributions of the signal, though some recourse to model calculations is needed to achieve this. At $10^{19}$ eV, $E_i$ (eV) comprises about 77% of the energy of the incoming cosmic ray. **It** can be found from an estimate of the total Cherenkov light, $Q_{tot}$, reaching the ground by integrating over all distances using an empirically-measured lateral distribution functions. The relationship [26] deduced is

$$\frac{E_i}{Q_{tot}} = (3.01 \pm 0.36) \times 10^4 (1 - \frac{X_{max}}{1700 \pm 270})$$

Thus measuring $E_i$ requires knowledge of $X_{max}$ (in g cm$^{-2}$) which is derived from the lateral distribution of Cherenkov light using shower models. The procedure is thus more dependent on knowledge of hadronic interactions than the approach using fluorescence radiation.

The work initiated by Chudakov continues to be a key supplementary feature of the scintillator array at Yakutsk used to study primaries >$10^{18}$ eV, while at the Tunka array near Lake Baikal an all-air Cherenkov system has been developed to characterize the properties of showers in the region of the knee near $10^{15}$ eV. In the latest implementation an area of 0.1 km$^2$ has been instrumented.

---

[4]Trevor Weekes claims that John Jelley found the night-time work at Harwell was so interrupted by bad weather that he asked the Harwell storeman if he could have a bed. The storeman is said to have replied "Certainly Dr Jelley. Would that be a single or a double bed?" Physicists in the UK could get almost anything in the days post-World War II.

Although more Cherenkov photons are produced per meter of electron track than by the fluorescence process, the light is directional and the fall-off with distance from the shower axis is similar to that of electrons. Coupled with the need for clear moonless nights for observation, this means that it is not an ideal tool for a large shower array. Furthermore shielding the photomultipliers from sunlight during daylight hours is a non-trivial exercise when one has many detectors. An alternative exploitation of Cherenkov radiation to detect large showers was proposed by Kieda [27] who suggested using solar cells as detectors, demonstrating that it might be possible to detect light under conditions of high background illumination. So far this technique has not been developed.

Chudakov [28] suggested that Cherenkov light produced in showers and reflected from snow to a sensor on a balloon or a plane might offer a way of achieving the vast collecting arrays needed to study the highest energy cosmic rays. The idea was tested by the Torino group who observed showers both simultaneously with scintillators and Cherenkov light detectors. The latter were pointed at the snow on the Plateau Glacier at the Testa Grigia station at 3500 m a.s.l. [29].

The potential of using Cherenkov radiation to cross-calibrate estimates of shower energies made at different arrays was pointed out by K. E. Turver and K. J. Orford who proposed transporting a single-photomultiplier detector from array to array and measure the signal at a reference distance. Attempts to carry out a cross-calibration between Volcano Ranch and Haverah Park in 1974 failed for a variety of reasons.

It worth noting that efforts to detect fluorescence radiation are hampered by the presence of the Cherenkov radiation produced by shower particles. Some of this radiation is scattered by molecules and aerosols into the photomultipliers of fluorescence telescopes giving a background for which allowance has to be made.

## 4 Use of water-Cherenkov detectors in the study of extensive air showers

Arrays of water-Cherenkov detectors have been used very successfully at Haverah Park and at the Auger Observatory. Credit for this development goes to N.A. Porter[5] who, while a member of the team working with the Geiger counter array at Harwell, became the first to succeed in preventing bacterial growth in unfiltered water long enough to realise a stable detector [30]. This was achieved by enclosing the water in a cubical, sealed, box of 'Darvic', a material then manufactured in the UK by ICI. Darvic was chosen primarily for its white diffusive surface. The depth of water was 92 cm. One of several advantages of a water-Cherenkov detector is that enables the energy flow in the shower to be measured. At 500 m from the axis of an event produced by a primary particle of ~ $10^{18}$ eV about half of the energy flow is carried by electrons and photons of mean energy ~ 10 MeV and these are very largely absorbed in the water [31]. Porter's detector can be seen as the prototype of those that were used at Haverah Park between 1967 and 1987) and at the Pierre Auger Observatory [32] which began operating in 2004. Indeed there has been remarkably little advance from Porter's design in which the photomultiplier looked downwards into the water.

When the Harwell array was closed in the mid-1950s to make space for the construction of the Culham Fusion Laboratories, it was decided, under the strong influence of Blackett, that work on extensive air-showers should continue in the United Kingdom but be supported and developed within the university environment. This led to the construction of the Haverah Park array under the leadership of J.G. Wilson, following prototype work at Silwood Park near London directed by H.R. Allan [33]. The depth of water in each 60 cm$^2$ detector of Allan's array was only 20 cm. Again the material in contact with the water and reflecting the light was chosen to be Darvic, on Porter's recommendation. At Haverah Park [34] the water depth was increased to 1.2 m and it was later shown by C.L. Pryke, during the design study undertaken for the Pierre Auger Observatory, that the number of photoelectrons detected in the standard geometry with the photomultiplier centred in the top surface of the tank has a broad maximum that peaks at 1.2 m. The author is unaware as to how the depth of water at Haverah Park was chosen and certainly no detailed Monte Carlo calculations were carried out. The depth may have been chosen through a combination of sound physical principles and a pragmatic approach to what steel sheet was available for the manufacture of the tanks which were lined on all sides with sheets of Darvic[6].

---

[5]Neil Porter was also the first person to report the detection of Cherenkov light produced by cosmic ray particles crossing the vitreous humour of the eye. A letter was published in the Psychology section of Nature (F.J. D'Arcy and N.A. Porter, Nature 196 (1962) 1013).

When the Auger Observatory was conceived, a number of possible devices were examined for the detectors of the surface array including RPCs, radio antenna, scintillators and water-Cherenkov detectors. The latter were chosen because of the capability of measuring energy flow and also because the depth adopted, 1.2 m as at Haverah Park, meant that showers of extremely large zenith angles could be measured with an accuracy that gave the potential to have whole sky coverage with only two observatories at intermediate latitudes and of studying showers at extremely large zenith angles where neutrino detection was thought to be possible.

An air-shower array (IceTop), constructed at the South Pole, will eventually consist of 80 pairs of tanks spread over 1 km$^2$ in which Cherenkov radiation produced in ice is detected. This array [35] will survey the energy range through the knee region at $10^{15}$ eV and to beyond $10^{18}$ eV. It can be operated independently or in conjunction with the giant neutrino detector, IceCube, (also a Cherenkov detector) which has just been completed.

The air-shower community owes P.A. Cherenkov a significant debt: detectors based on his discovery are likely to provide rich science for many years to come.

## 5 Acknowledgements

I would like to thank Antonio Insolia for making it possible for me to attend the 2010 CRIS meeting. A number of (some times conflicting) insights into the issue of the relationship that Cherenkov had with fellow scientists in the USSR have come to me through conversations and correspondence with Venya Berezinsky, Sasha Lidvansky, Anatoly Petrukhin and Christian Spiering: the interpretation of these inputs is mine. Some information about Harwell days originated with Giorgio Palumbo and Trevor Weekes, both of whom worked closely with John Jelley.

---

[6]Darvic had been developed by ICI for the manufacture of sandwich boxes and so a material containing a bacteriological inhibiter was invented. The author is indebted to Professor I.M. Ward (University of Leeds, but formerly of ICI) for this insight which came many years after the closure of the Haverah Park array where the folklore was that zinc in the galvanizing of the steel tanks inhibited growth of bacteria.